\documentclass[a4paper, amsfonts, amssymb, amsmath, reprint, showkeys, nofootinbib, superscriptaddress, twoside]{revtex4-1}
\usepackage[english]{babel}
\usepackage[utf8]{inputenc}
\usepackage{CJK}
\usepackage[colorinlistoftodos, color=green!40, prependcaption]{todonotes}
\usepackage{amsthm}
\usepackage{mathtools}
\usepackage{physics}
\usepackage{xcolor}
\usepackage{graphicx}
\usepackage[left=23mm,right=13mm,top=35mm,columnsep=15pt]{geometry} 
\usepackage{adjustbox}
\usepackage{placeins}
\usepackage[T1]{fontenc}
\usepackage{lipsum}
\usepackage{csquotes}
\usepackage[pdftex, pdftitle={Article}, pdfauthor={Author}]{hyperref} 
\begin{document}
\begin{CJK}{UTF8}{gbsn}
\title{Study of jet substructure modification by differential girth at the LHC}

\author{Si-Yu Tang(汤思宇)}
    \affiliation{School of Mathematical and Physical Sciences, Wuhan Textile University, Wuhan 430200, China}
    \affiliation{Hubei Key Laboratory of Digital Textile Equipment, Wuhan Textile University, Wuhan 430200, China}
\author{Wei Zhang(张伟)} 
    \affiliation{School of Electronic and Electrical Engineering, Wuhan Textile University, Wuhan 430200, China}
\author{Wen-Di Deng(邓文迪)} 
    \affiliation{School of Electronic and Electrical Engineering, Wuhan Textile University, Wuhan 430200, China}
\author{Cheng-De Tian(田成得)} 
    \affiliation{School of Electronic and Electrical Engineering, Wuhan Textile University, Wuhan 430200, China}
\author{Fan Yang(杨帆)}
    \affiliation{School of Mathematical and Physical Sciences, Wuhan Textile University, Wuhan 430200, China}
\author{Ren-Zhuo Wan(万仁卓)}
    \email[Correspondence email address: ]{wanrz@wtu.edu.cn}
    \affiliation{Hubei Key Laboratory of Digital Textile Equipment, Wuhan Textile University, Wuhan 430200, China}
    \affiliation{School of Electronic and Electrical Engineering, Wuhan Textile University, Wuhan 430200, China}
\date{\today} 

\begin{abstract}
Jet substructure observables are crucial for exploring the effects of the hot, dense medium and differentiating between quark and gluon jets. In this paper, we investigate the modification of jet shape by calculating the differential girth at LHC energies, ranging from pp to Pb--Pb collisions. The differential girth distribution exhibits a wave-like pattern, which offers a better explanation for the radial spreading of jet energy and large-angle radiation effects. We fit the differential girth distribution with a sine wave function to obtain the amplitude, angular frequency, initial phase, and offset distance parameters as a function of jet radius. The results clearly differentiate between pp and heavy-ion collisions, with varying dependencies on centrality and collision energy. Our findings shed new light on the understanding of radial jet energy loss in the QGP (quark-gluon plasma) medium, and provide additional potential observables for measuring jet shape at the LHC in the future.
\end{abstract}


\maketitle
\section{Introduction}
Probing the properties of the hot-dense QCD matter, quark-gluon plasma (QGP), created in ultra-relativistic heavy-ion collisions is one of main goals for high energy physics. Number of physics observables, such as particle multiplicity, strangeness enhancement, hadron correlation, anisotropic flow and jet quenching etc., are proposed to explore the underlying mechanism of particle production in nuclear environment. Among these observables, the jets, which originate from initial hard scatterings with large momentum transfer, are sensitive to the whole evolution of QGP medium. The modification of the production of jets in heavy-ion collisions compared to binary-scale pp collisions, so-called jet quenching, reflects the energy reduction of hard partons in hot-dense QGP medium. However, great challenges arise from the description of medium recoil, thermalization and induced gluon radiation in phenomena, as well as the separation in experiment. In order to ultimately provide a way to address these physics concerns, the study of the internal behaviors of particles within charged jets, i.e. jet substructure, is performed. It can provide us an insight to interplay the perturbative and non-perturbative physics, as well as to further probe the jet formation and evolution. Recently several observables related to jet substructure were measured at the LHC with high resolution, such as the jet mass, jet momentum dispersion, jet angularity~\cite{ALICE:2018dxf}, groomed jet momentum splitting fraction and radius, as well as $N$-subjettiness~\cite{ALICE:2021vrw}, fragmentation functions and so.on. In the measurements of fragmentation functions of charged hadrons for jets~\cite{PhysRevC.90.024908,PhysRevLett.123.042001,ALICE:2022vsz}, a significant modification of the fragmentation function is observed in most central Pb--Pb collisions compared to pp collisions, and the difference in fragmentation between light quarks and gluons is also studied in order to investigate the parton color-charge dependence of jet quenching in the QGP medium. In addition, the sub-jet fragmentation measured in~\cite{ALICE:2021njq} is regarded as a complement to the longitudinal momentum fraction ($z$) of hadrons in jets, especially for probing high $z$. No significant modifications of the $z_{r}$ (the fraction of transverse momentum ($\it{p}_{\rm T}$) carried by the subjet) distribution is observed from central Pb--Pb collisions to pp collisions, while a hardening effect is also shown at intermediate $z_{r}$ due to the larger relative suppression of gluon jets to quark jets in QGP medium. Interestingly such effect turns over as $z_{r} \rightarrow 1$, indicating the competing effects from medium-induced soft radiations. On the other hand, the measurements of jet angularities in heavy-ion and pp collisions~\cite{ALICE:2018dxf} also provides us the straight information of the medium modification of the jet shape, which indicates that the in-medium fragmentation is harder and more collimated compared to the vacumn. Furthermore, with the development of Jet grooming techniques~\cite{PhysRevLett.119.112301} in recent years, the groomed jet observables, such as groomed jet momentum splitting fraction ($Z_{g}$) and jet radius ($\theta_{g}$), are also measured by ALICE~\cite{PhysRevLett.128.102001}. A narrower $\theta_{g}$ distribution in Pb--Pb collisions compared to pp collisions is clearly observed and no significant modification of the $z_{g}$ distribution is shown. It provides the direct evidence to identify the modification of the angular structure of jets in the quark-gluon plasma. However these behaviours of jet shape from pp to heavy-ion collision systems can be well described by the theoretical models~\cite{PhysRevLett.119.112301,Mehtar-Tani:2016aco} with different, even opposite jet quenching mechanisms, which leave us many unknowns in theory including the roles of coherent and incoherent energy loss~\cite{Zapp:2011ya}, medium response~\cite{KunnawalkamElayavalli:2017hxo}, and the space-time picture of the parton shower and so.on. In order to disentangle the physics explanatory behind these models, new observables are strongly encouraged to be proposed. In this paper, we further investigate the observable differential girth proposed in our previous work~\cite{Wan:2018zpq} with JEWEL simulation.

\section{Differential Jet Girth}
Among a plenty of jet-shape observables, the jet girth $g$, also called first radial moment or angularity, is proposed to explore the radial energy distribution inside jets. It is defined as the distance between the jet constituent and the jet axis weighted by its transverse momentum ($\it{p}_{\rm T}$):
\begin{equation}
g=\sum_{i\in jet}\frac{p_{\rm T,i}}{p_{\rm T,jet}}|\Delta R_{i,\rm jet}|.
\label{def: jet girth}
\end{equation}
Here the $p_{\rm T,i}$ represents $\it{p}_{\rm T}$ of $i$-th constituent inside the jets with $p_{\rm T,jet}$ and the $\Delta R_{i,\rm jet}$ is the distance between the $i$-th constituent and the jet axis in ($\eta,\varphi$) space, where $\eta$ and $\varphi$ is the pseudorapidity and azimuthal angle, respectively. This observable has been measured at the LHC for small-radius jets~\cite{ALICE:2018dxf} in both pp and Pb--Pb collisions. The girth distribution in heavy-ion collisions shifted to the lower value compared to the pp collisions, which indicates more collimated and harder jet core at the same reconstructed jet energy. 

In order to investigate the evolution of the energy distribution inside the jets, especially for the jets with small radii $r$ ($r<0.3$) which carry most high $p_{\rm T}$ constituents, the differential jet girth, $D_{\rm girth}$ was proposed~\cite{Wan:2018zpq}. It is defined as the girth distribution inside an annulus of inner radius $r$ and outer radius $r+\Delta r$ around the jet axis:
\begin{equation}
D_{\rm girth}(r+\Delta r) = g(r+\Delta r) - g(r)
\label{def: jet girth}
\end{equation}
Compared to the traditional observable girth $g$, the differential girth $D_{\rm girth}$ is more sensitive to the dynamic energy flow of jets along the radial direction from jet axis. It was observed that the differential girth in pp collisions is different from that in Pb--Pb collisions for fixed jet $p_{\rm T}$ and radius with JEWEL simulation~\cite{Wan:2018zpq}, which is due to the parton energy loss in hot dense medium created in heavy-ion collisions. In order to quantify the effects of jet-quenching on the jet shape, we test the dependence of differential girth on collision energy in pp collisions and centrality selections in Pb--Pb collisions with JEWEL simulation in this paper, then further extract the evolution parameters to describe the propagation of jet components in vacuum and medium, respectively.

\section{JEWEL}
 The JEWEL (Jet Evolution With Energy Loss) is a common Monte Carlo event generator simulating QCD jet evolution in heavy-ion collisions~\cite{Zapp:2013vla}. It describes the global properties of jets in a perturbative approach at the LHC energy. A virtuality-ordered parton is implemented in JETWEL with the evolution parameter, i.e. parton virtuality, $Q^{2}$. In the current version of JEWEL (2.2.0), a sufficiently hard scattering in the medium modifies the projectile virtuality, which affects the shower evolution. The Landau-Migdal-Pomeranchuk (LPM) effect~\cite{Landau:1953um} is included by generalizing the probabilistic formulation, thus a parton cannot undergo a splitting instantaneously, but only with a finite formation time. On the other hand, a variant of the Bjorken model~\cite{KunnawalkamElayavalli:2017hxo} which describes the boost-invariant longitudinal expansion of an ideal QGP is set. The shower initiated by the hard scattered parton interacts with the background partons and loses energy according to elastic and radiative processes. For the treatment of medium response, there are two set of operational modes in JEWEL: recoil ON and OFF. In the recoil OFF mode, the recoiled medium partons do not participate in further processes towards hadronization. While in the recoil ON version, the effects of recoiling partons are propagated into the final observables by means of inserting them into the strings connecting the parton shower. The results obtained from recoil ON version can improves the description of the jet shape observables in JEWEL, however faces potential challenges from the background subtraction. In this study, the jets are reconstructed by all final detectable particles with the anti-$k_{\mathrm{T}}$ jet finding algorithm~\cite{Cacciari:2008gp} using the Fastjet package~\cite{Cacciari:2011ma} at midrapidity $|\eta|<2$. The recoil mode is set ON, and the contribution of medium fragments from the parton energy is properly removed by the constituent subtraction method.

\section{Results and discussion}
\begin{figure}[!hbt]
  \begin{center}
  \includegraphics[width=.99\columnwidth]{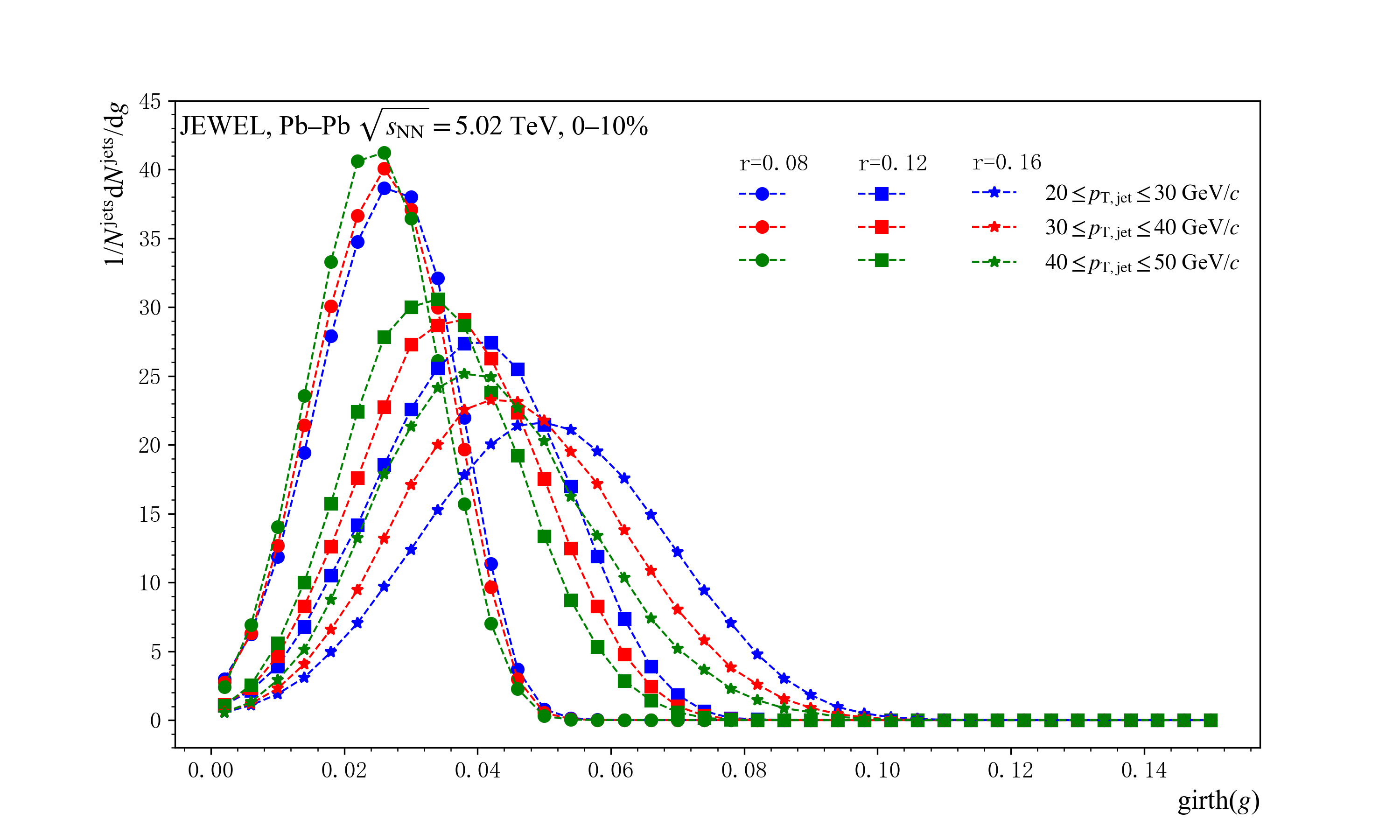}
  \includegraphics[width=.99\columnwidth]{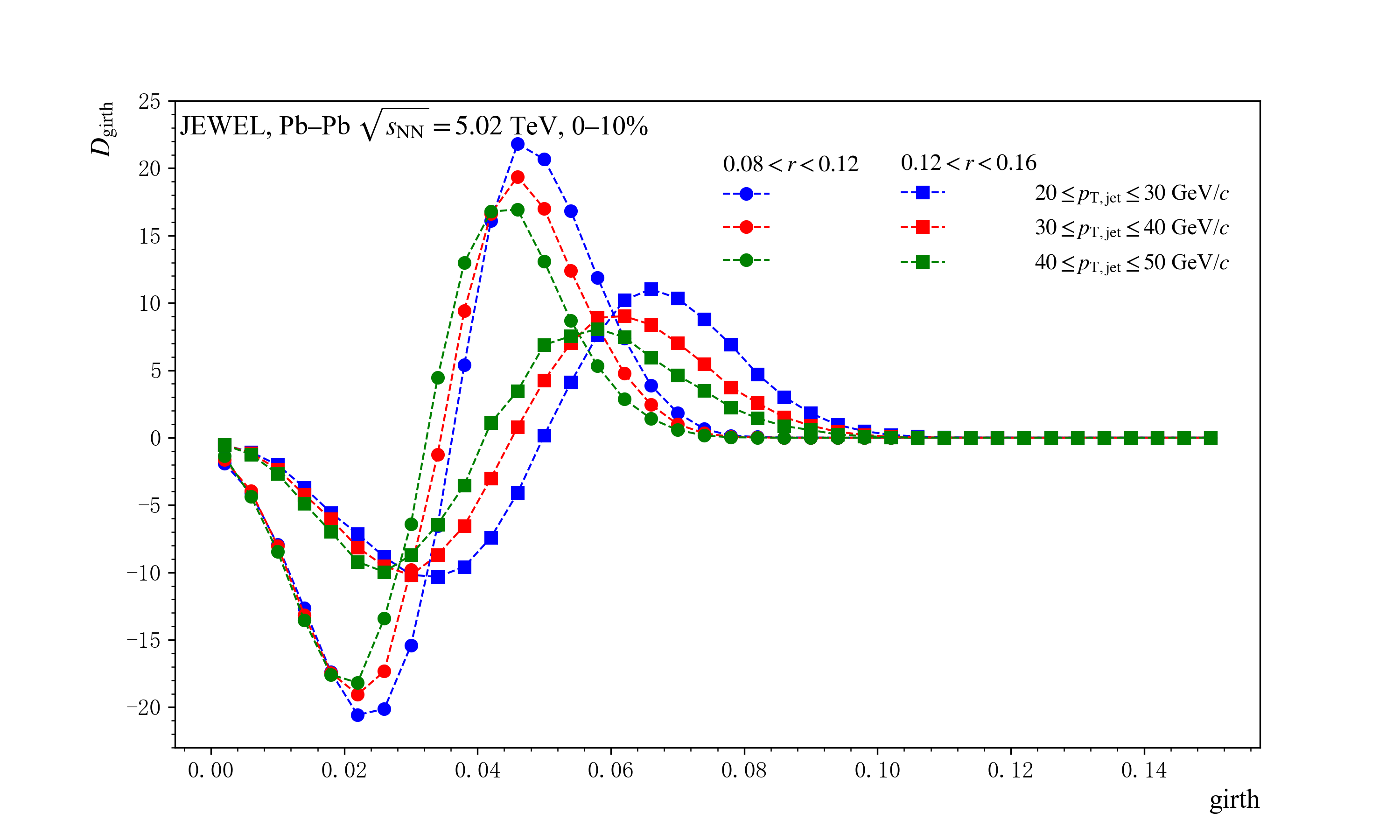}
  \caption{The girth (top) and differential girth (bottom) distribution with different jet $p_{\rm T}$ selections in 0--10\% Pb--Pb collisions at $\sqrt{s_{\rm NN}}$ = 5.02 TeV simulated by JEWEL.}
  \label{Fig: Girth in PbPb}
\end{center}
\end{figure}
Figure~\ref{Fig: Girth in PbPb} (top) shows the distribution of girth in Pb--Pb collisions at $\sqrt{s_{\rm NN}}$ = 5.02 TeV with the selections of jet $p_{\rm T}$ and radius, obtained from JEWEL simulation. One can see that the trend of girth distribution is similar to the data, and the peak values shift to higher values with the increasing $p_{\rm T}$ and decreasing radius of jets. It is what we expected since the jets with higher $p_{\rm T}$ are fragmented from the harder partons which can generate the harder jet core, and the medium induced modifications can also lead to the broadening of the jets with larger radius. Similarly, the distribution of differential girth, $D_{\rm girth}$ is shown in Fig~\ref{Fig: Girth in PbPb} (bottom). It is observed that the differential girth distribution behaves as a waveform, and becomes flat at large values. One can also see that this wave-like pattern propagates faster for the jets with smaller radius $r$ and larger $p_{\rm T}$, since they have smaller granularity and suffer larger jet energy loss in QGP medium. 

\begin{figure}[!hbt]
  \begin{center}
  \includegraphics[width=.99\columnwidth]{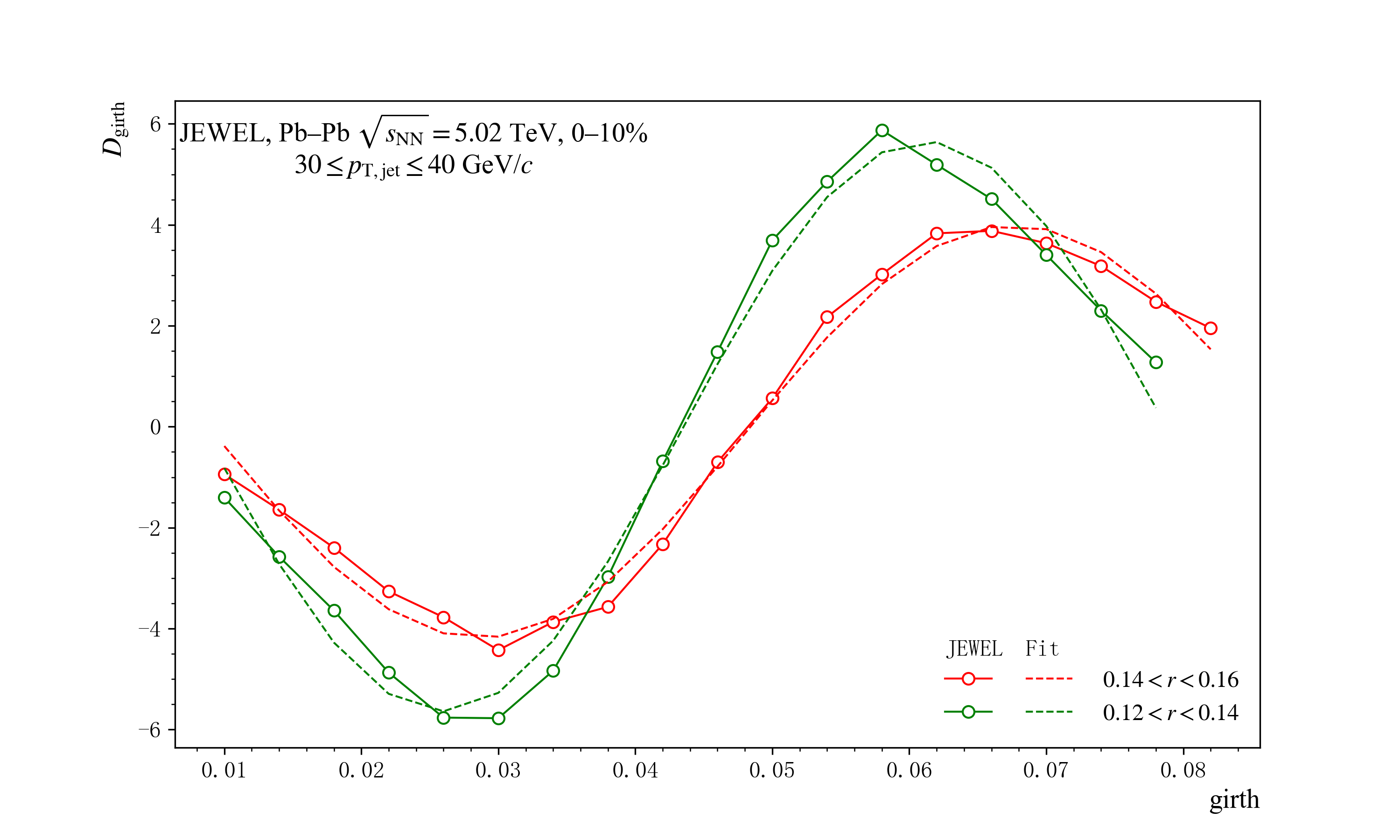}
  \caption{The fit of differential girth distribution in 0--10\% central Pb--Pb collisions at $\sqrt{s_{\rm NN}}$ = 5.02 TeV for $0.12<r<0.14$~cm and $0.14<r<0.16$~cm with the selection of jet $p_{\rm T}$ in 30$~<p_{\mathrm{T}}<~$40 GeV/$c$.}
  \label{Fig: Fit of diff Girth in PbPb}
\end{center}
\end{figure}

As introduced earlier, the characterization of magnitude of differential girth can help to understand the evolution of energy loss in QGP medium. We introduce the Sine wave function to fit the differential girth distribution shown in Fig~\ref{Fig: Girth in PbPb} (bottom). The function is defined as:
\begin{equation}
f(r+\Delta r) = \mathrm{A}\cdot\mathrm{sin}(\mathrm{B}\cdot r+\mathrm{C})+\mathrm{D}
\label{def: sin fit}
\end{equation}
where $r$ is the radius of the jet. The parameter A, B, C and D represent the amplitude, angular frequency, initial phase and offset distance of this Sine wave function, respectively. Figure~\ref{Fig: Fit of diff Girth in PbPb} shows an example of the fit of differential girth distribution in Pb--Pb collisions at $\sqrt{s_{\rm NN}}$ = 5.02 TeV obtained from JEWEL simulation. One can see that the fit works well and the parameters A, B, C and D can be extracted directly.

\begin{figure*}[!hbt]
  \begin{center}
  \includegraphics[width=1.9\columnwidth]{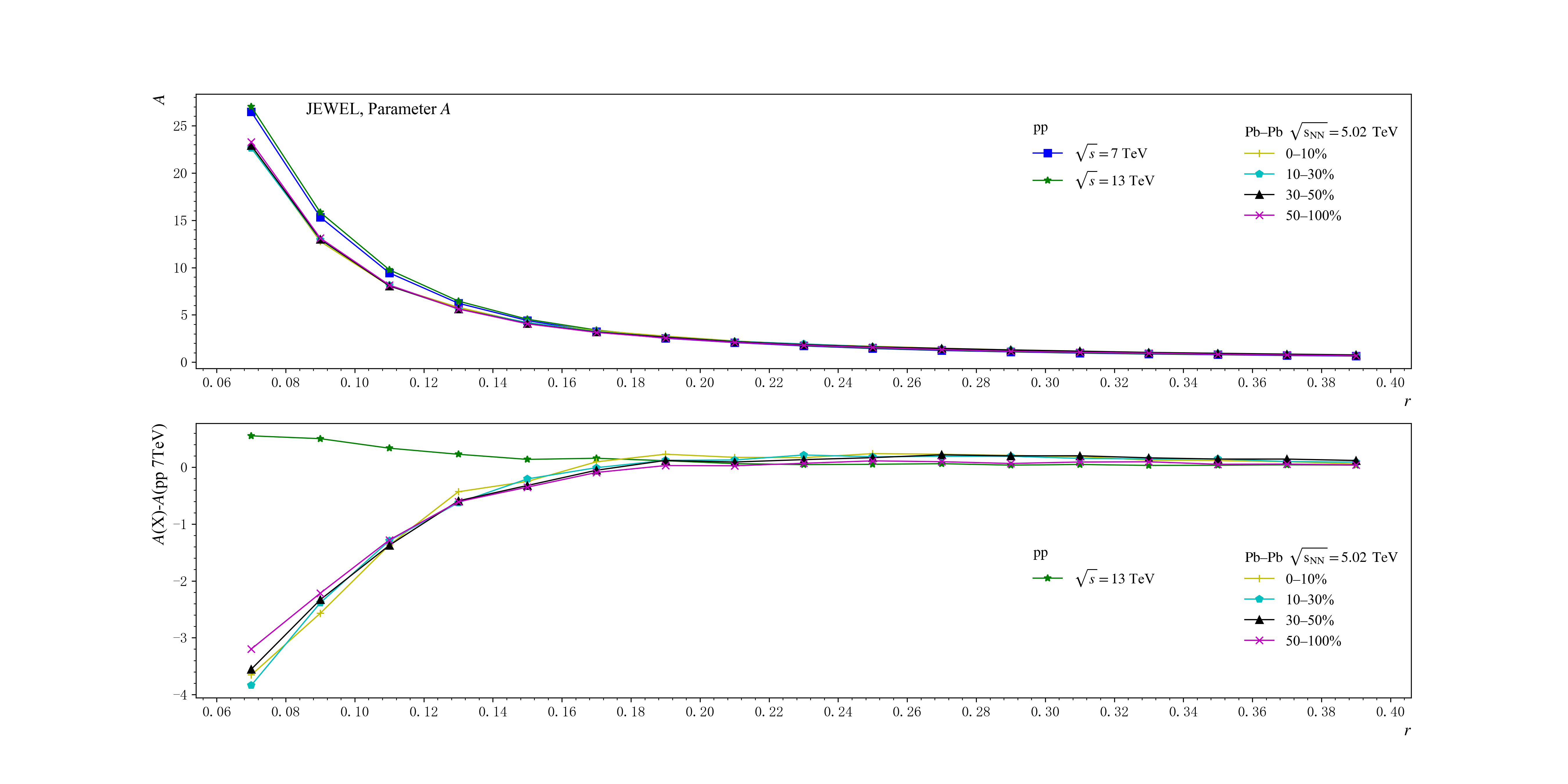}
  \caption{Top: the parameter A extracted in differential girth distribution as a function of jet radius, obtained in pp collisions at $\sqrt{s}$ = 7 TeV, 13 TeV and Pb--Pb collisions at $\sqrt{s_{\rm NN}}$ = 5.02 TeV simulated by JEWEL. Different centrality selections are applied and the $p_{\rm T}$ of jets is chosen in 30$~<p_{\mathrm{T}}<~$40 GeV/$c$. Bottom: the deviation of parameter A obtained in pp collisions at $\sqrt{s}$ = 7 GeV and all other cases (labeled as "X" in y-axis).}
  \label{Fig: Parameter A}
\end{center}
\end{figure*}

In Fig~\ref{Fig: Parameter A}(top), the extracted parameter A is presented as a function of jet radius for pp collisions at $\sqrt{s}$ = 7 and 13 TeV and Pb--Pb collisions at $\sqrt{s_{\rm NN}}$ = 5.02 TeV, as well as for different centrality selections in Pb--Pb collisions. The bottom panel of Fig~\ref{Fig: Parameter A} displays the deviation between the results obtained in pp collisions at $\sqrt{s}$ = 7 TeV and other cases. The parameter A obtained in pp collisions at $r<0.16$ is significantly higher than that obtained in Pb--Pb collisions, indicating that the magnitude of the energy decrement along the radial direction is smaller in Pb--Pb collisions than in pp collisions, as expected. This finding can be attributed to the fact that the jet core in Pb--Pb collisions is more collimated, which means that the energy of the jet is concentrated in a smaller area around the jet axis, resulting in less energy being lost to the medium. In contrast, the jet in pp collisions has a wider core, which leads to a larger energy loss along the radial direction. In addition, there is no significant dependence on centrality or collision energy in the differential girth observed in both Pb--Pb and pp collisions.

\begin{figure*}[!hbt]
  \begin{center}
  \includegraphics[width=2.0\columnwidth]{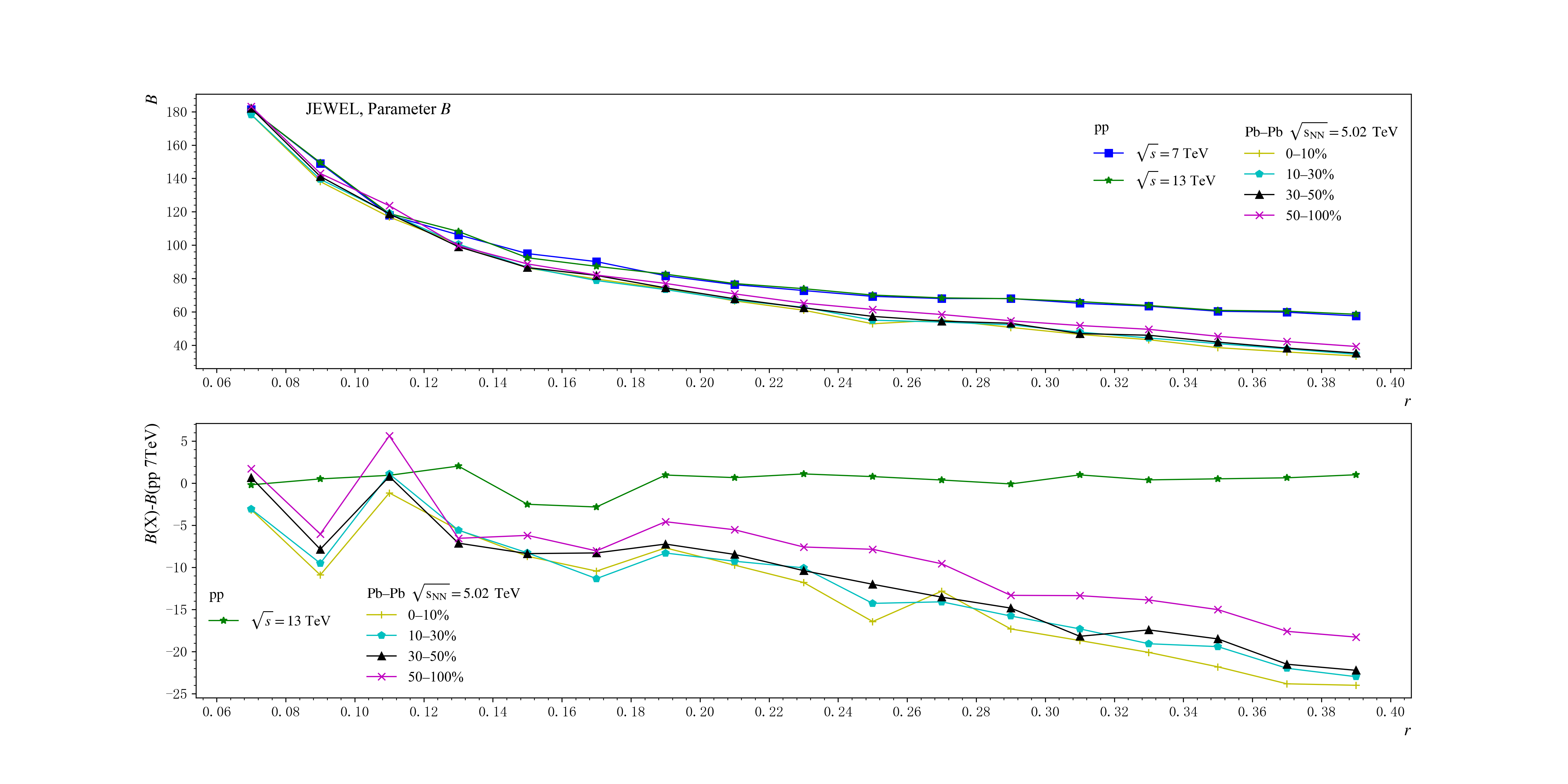}
  \caption{Top: the parameter B extracted in differential girth distribution as a function of jet radius, obtained in pp collisions at $\sqrt{s}$ = 7 TeV, 13 TeV and Pb--Pb collisions at $\sqrt{s_{\rm NN}}$ = 5.02 TeV simulated by JEWEL. Different centrality selections are applied and the $p_{\rm T}$ of jets is chosen in 30$~<p_{\mathrm{T}}<~$40 GeV/$c$. Bottom: the deviation of parameter B obtained in pp collisions at $\sqrt{s}$ = 7 TeV and all other cases (labeled as "X" in y-axis).}
  \label{Fig: Parameter B}
\end{center}
\end{figure*}

Figure~\ref{Fig: Parameter B} (top) presents the extracted parameter B as a function of the radius of jets in pp collisions and Pb--Pb collisions at different collision energy. The top panel of Figure~\ref{Fig: Parameter B} shows the extracted parameter B as a function of jet radius for pp and Pb--Pb collisions at various collision energies. As the jet radius increases, the differential girth decreases in both pp and Pb--Pb collisions. However, the reduction rate in Pb--Pb collisions is larger than that in pp collisions. The bottom panel of Fig~\ref{Fig: Parameter B} displays the deviation of the differential girth obtained in pp collisions at $\sqrt{s}$ = 7 TeV from that in pp collisions at $\sqrt{s}$ = 13 TeV and Pb--Pb collisions. A clear distinction between the results in pp and Pb--Pb collisions is observed, particularly for larger radii. The parameter B is greater in pp collisions than in Pb--Pb collisions, indicating that jet components travel faster in a vacuum than in a medium along the radial direction. Moreover, the parameter B is systematically lower in smaller centrality classes, since more central collisions produce a greater density of particles, resulting in stronger interactions between the jet and the medium. This observable has the potential to provide additional insights into the tuning of jet quenching mechanisms, as confirmed by future experimental findings.

\begin{figure*}[!hbt]
  \begin{center}
  \includegraphics[width=2.0\columnwidth]{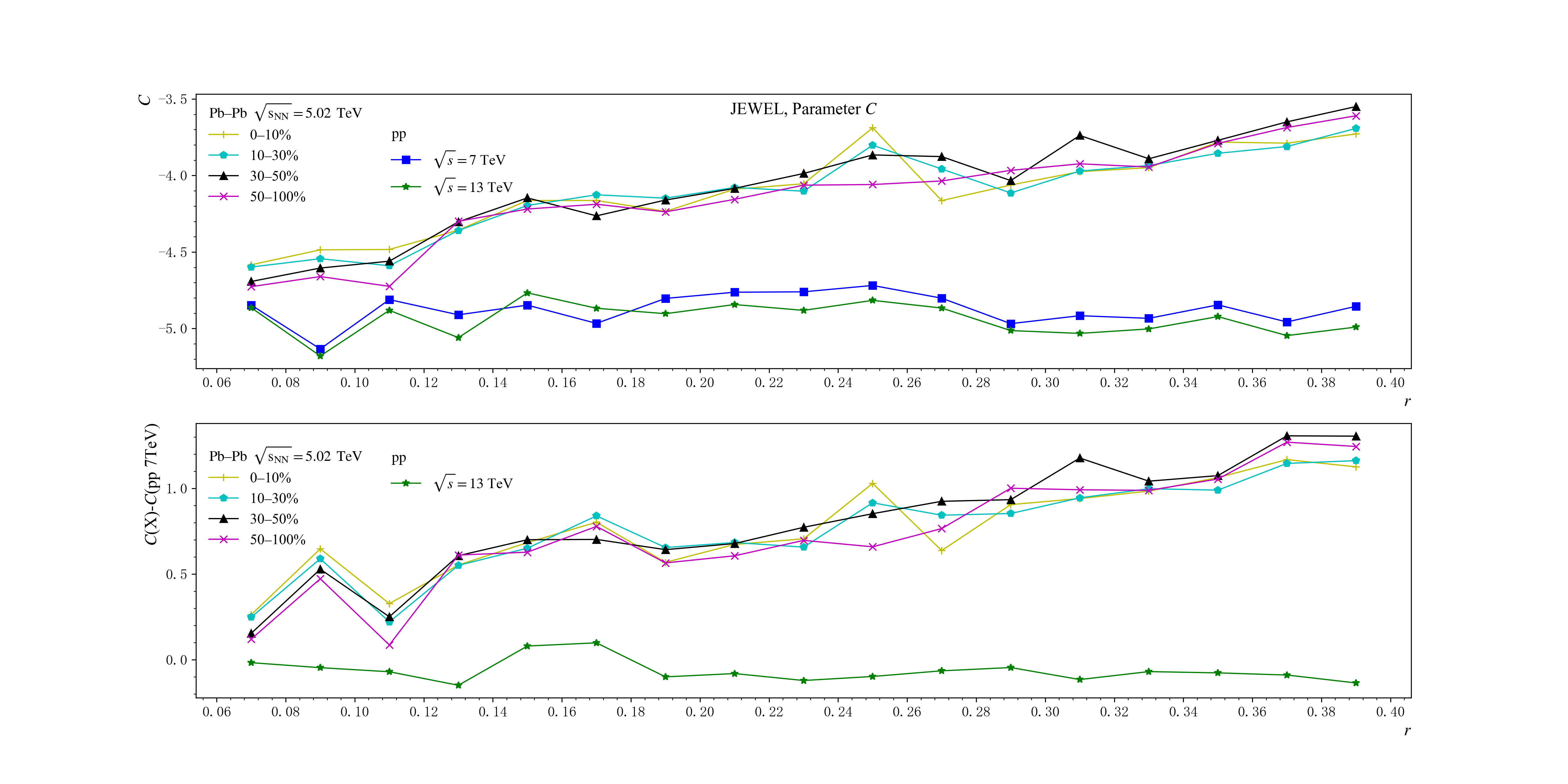}
  \caption{Top: the parameter C extracted in differential girth distribution as a function of jet radius, obtained in pp collisions at $\sqrt{s}$ = 7 TeV, 13 TeV and Pb--Pb collisions at $\sqrt{s_{\rm NN}}$ = 5.02 TeV simulated by JEWEL. Different centrality selections are applied and the $p_{\rm T}$ of jets is chosen in 30$~<p_{\mathrm{T}}<~$40 GeV/$c$. Bottom: the deviation of parameter C obtained in pp collisions at $\sqrt{s}$ = 7 GeV and all other cases (labeled as "X" in y-axis).}
  \label{Fig: Parameter C}
\end{center}
\end{figure*}

\begin{figure*}[!hbt]
  \begin{center}
  \includegraphics[width=2.0\columnwidth]{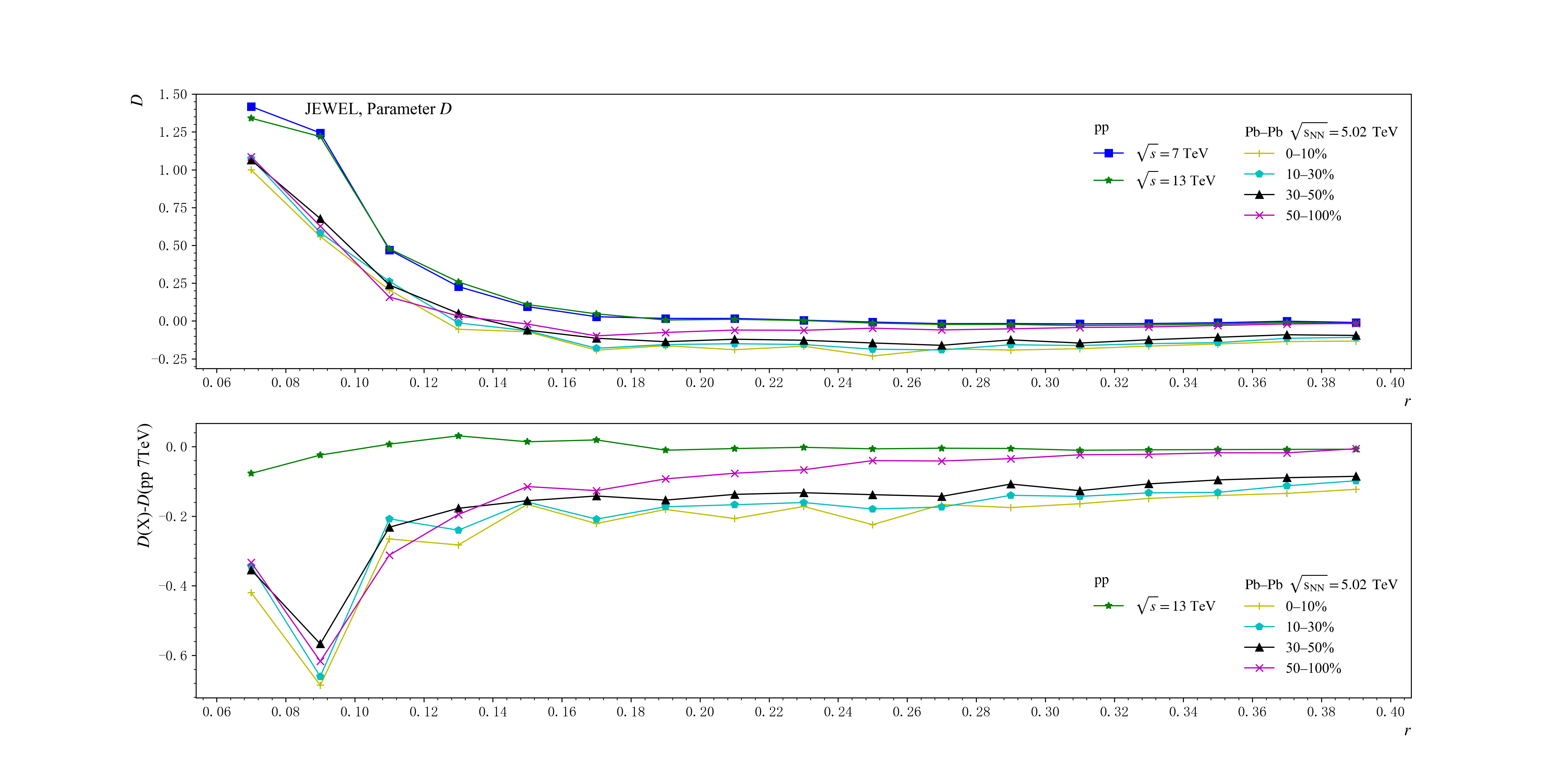}
  \caption{Top: the parameter D extracted in differential girth distribution as a function of jet radius, obtained in pp collisions at $\sqrt{s}$ = 7 TeV, 13 TeV and Pb--Pb collisions at $\sqrt{s_{\rm NN}}$ = 5.02 TeV simulated by JEWEL. Different centrality selections are applied and the $p_{\rm T}$ of jets is chosen in 30$~<p_{\mathrm{T}}<~$40 GeV/$c$. Bottom: the deviation of parameter D obtained in pp collisions at $\sqrt{s}$ = 7 TeV and all other cases(labeled as "X" in y-axis).}
  \label{Fig: Parameter D}
\end{center}
\end{figure*}

Figures~\ref{Fig: Parameter C} and Fig.~\ref{Fig: Parameter D} display the extracted parameters C and D, respectively, as a function of the radius of jets in pp collisions at $\sqrt{s}$ = 7 and 13 TeV and Pb--Pb collisions at $\sqrt{s_{\rm NN}}$ = 5.02 TeV, along with the results for different centrality selections in Pb--Pb collisions. Deviations from the results in pp collisions at $\sqrt{s}$ = 7 TeV are also shown.  In Pb--Pb collisions, the parameter C increases with increasing jet radius, indicating a harder initial energy distribution. This observation is consistent with the experimental observation that the jets in Pb--Pb collisions are more collimated than those in pp collisions. Moreover, a clear separation between the results in pp and Pb--Pb collisions is observed, especially for high-radius jets. Conversely, the parameter C extracted from pp collisions does not exhibit a significant dependence on the jet radius. Regarding the parameter D, a clear separation between pp and (semi-)central Pb--Pb collisions is observed, with the latter having a lower value. The smaller offset distance of the Sine wave function in Pb--Pb collisions compared to pp collisions suggests that there is a large-angle radiation in Pb--Pb collisions due to the energy loss of the jets in the medium. Interestingly, the parameter D in peripheral 50--100\% Pb--Pb collisions is consistent with that in other Pb--Pb cases with small jet radii, whereas it is close to that in pp collisions with large jet radii. It may suggest that there is a smooth transition for the modification of jet shape between large and small collision systems. These findings provide an alternative way to characterize the magnitude of jet quenching in small collision systems, complementing the traditional hadron-jet correlations measurements~\cite{ALICE:2017svf}.

\section{Conclusions}
This paper presents a detailed study of the differential girth distribution in pp and heavy-ion collisions at LHC energies using JEWEL simulation. The differential girth distribution exhibits a wave-like pattern, and the amplitude, angular frequency, initial phase, and offset distance parameters as a function of jet radius are obtained by fitting the differential girth distribution with a sine wave function. By revealing the wave-like pattern of the differential girth distribution and characterizing its parameters, our analysis offers new insights into the radial energy loss of jets in the QGP medium. Moreover, the obtained parameters provide a valuable tool to discern the jet shape evolution between large and small collision systems, which may serve as a potential observable for future jet measurements at the LHC.

\section{Acknowledgement}
We thank X.~Zhang, D.~Zhou for discussions. This work was supported by Natural Science Foundation of Hubei Provincial Education Department under Grant (Q20131603) and Key Laboratory of Quark and Lepton Physics (MOE) in Central
China Normal University under Grant (QLPL2022P01, QLPL202106).

\bibliographystyle{apsrev4-2} 
\bibliography{reference}





\end{CJK}
\end{document}